\documentclass[useAMS,usenatbib]{mn2e}

\usepackage{graphicx}
\usepackage{amssymb}
\usepackage{pstricks}
\usepackage{longtable,lscape}

\def \be {\begin{equation}}
\def \ee {\end{equation}}

\def \be {\begin{equation}}

\def \ee {\end{equation}}

\title{Early Dynamical Instabilities in the Giant Planet Systems}

\author[Elena Lega, A. Morbidelli and D. Nesvorn\'y] { 
E. Lega$^{1}$, \thanks{E-mail: elena@oca.eu (EL); morby@oca.eu (AM); davidn@boulder.swri.edu (DN)}, 
A. Morbidelli$^{1}$  and D. Nesvorn\'y$^{2}$\\
$^{1}$ Universit\'e de Nice Sophia Antipolis, CNRS UMR 7293,  Observatoire de la
C\^ote d'Azur,   Bv. de l'Observatoire, \\B.P.~4229,  06304 Nice cedex 4, 
France.\\
$^{2}$ Southwest Research Institute, Department of Space Studies, 1050 Walnut St., 
Suite 300, Boulder, Colorado 80302 \\
}

\begin{document}

\date{Accepted .... Received ....; in original form ....}

\pagerange{\pageref{firstpage}--\pageref{lastpage}} \pubyear{2012}

\maketitle
\label{firstpage}

\begin{abstract}
The observed wide eccentricity distribution of extrasolar giant planets 
is thought to be the result of dynamical instabilities and gravitational
scattering among planets. Previously, it has been assumed that the orbits 
in giant planet systems become gravitationally unstable after the gas 
nebula dispersal. It was not well understood, however, how these unstable 
conditions were established in the first place.   

In this work we numerically simulate the evolution of systems of three planets 
as the planets sequentially grow to Jupiter's mass, and dynamically interact 
among themselves and with the gas disk. We use the hydro-dynamical code 
FARGO that we modified by implementing the $N$-body integrator SyMBA. The 
new code can handle  close encounters and collisions between planets. To 
test their stability, the planetary systems were followed with SyMBA for 
up to $10^8$ yr after the gas disk dispersal. 

We find that dynamics of the growing planets is complex, because migration and 
resonances raise their orbital eccentricities, and cause dynamical instabilities 
when gas is still around. If the dynamical instabilities occur early, planets 
can be removed by collisions and ejections, and the system rearranges into a 
new, more stable configuration. In this case, the planetary systems emerging 
from the gas disks are expected to be stable, and would need to be destabilized 
by other means (low-mass planets, planetesimal disks, etc.). Alternatively, for 
the giant planet system to be intrinsically unstable upon the gas disk dispersal, 
a special timing would be required with the growth of (at least some of) the 
giant planets having to occur near the end of the gas disk lifetime.
\end{abstract}

\section{Introduction}
The eccentricity distribution of the extrasolar giant planets is wide with 
orbits commonly having $e>0.3$. Such a wide distribution was unexpected based 
on our anticipation from the Solar System planets. Different mechanisms have 
been proposed to explain the high eccentricities values:  
trapping of planetary pairs in mean motion resonances (\citet{Lee02}),  
the Kozai cycles in binary systems (\citet{Mazeh97,Holman97}), 
stellar jets (\citet{Namouni05}), and gravitational scattering during 
global dynamical instabilities (\citet{WeidMarz96,RaFo96,LinIda97}). 
\par
This last mechanism has been investigated with extensive $N$-body simulations  
 (\citet{Chatterjee08,Raymond09,Juric08}). All these studies assumed that the planetary systems
emerging from the gas disks are intrinsically unstable, and the gravitational
interactions among planets cause instabilities after the gas disk dispersal.
The subsequent scattering encounters between planets lead to large orbital 
eccentricities, just as needed to explain the observations.
\par
A hydro-dynamical code has been recently used to study the planetary system 
instabilities in a low-density rapidly-dispersing disk (\citet{Moeckel12}). 
The initial orbits of the planets were assumed to be circular and 
close enough to each other to be unstable, without any mutual resonance 
relationship. It was found that the disk can stabilize some of the planetary 
systems by driving them into resonance rapidly. However, the systems that 
became unstable ended up behaving as in the gas-free simulations. 
\par
The investigations discussed above, including \citet{Moeckel12}, adopt similar 
choices of initial conditions with unstable and sometimes overlapping planetary 
orbits. In reality, the initial conditions of these studies should be informed 
from the previous stages of planet-disk interactions when the damping effects 
of gas were important.   
\par
The orbital dynamics of giant planets in a massive gas disk has been studied 
with a hydro-dynamical code by \citet{Marzari10}. They started their simulations 
with the fully-formed giant planets and ignored the previous stage during which 
the giant planets had grown by gas accretion onto their cores. 
\par 
Here we report the result of the first effort to investigate the dynamical evolution of planets from their growth phase from cores to aftermath 
of the gas disk dispersal.
{ Our simulation set up is different from those of previous works
and is defined according to the following rationale. Planetary cores are expected to form by oligarchic growth (\citet{KoIda98}) with orbital separations of about 10 mutual Hill radii. However, during and after their formation they migrate in the disc due to their gravitational interactions with the gas. According to the Pollack et al. model (\citet{Pollack96}) the cores can spend a few millions years in
the disc  before accreting gas in a runaway fashion and become   giant planets.
In this time they can substantially modify their orbits and reach a new equilibrium configuration. While it was thought in the past that cores continuously migrate toward the central star (\citet{Ward97}), it is now known that in a disc with realistic heat diffusion they migrate towards an orbital radius where migration is cancelled (\citet{PaardeMelle06,KBK09,Lyraetal10}). This no-migration radius acts as a {\it planet trap}. If multiple cores are present they are expected to reach a resonant non migrating configuration near the trap (\citet{MorbyCrida08}).
In this configuration the cores can be much closer to each other than their initial 10 Hill radii separation which may lead to very strong instabilities when the planets grow to Jupiter mass.\par
With this kind of dynamical evolution in mind, in our hydrodynamical  simulations we set up a planet trap and let a system of three embryos of 10 Earth masses to evolve until they reach a stable resonant configuration.}
Then, we track the evolution of  the systems 
 as each of the three  planets grows in sequence  to one Jupiter 
mass. Finally, we slowly remove the gas from the disc and
 follow the evolution of the systems  up to $10^8$ years after the gas 
dispersal. We use the hydro-dynamical code FARGO (\citet{Masset00}) modified in 
\citet{MorbyNesvo12} by implementing the $N$-body integrator SyMBA (\citet{Duncan98}) 
\footnote{We use {\tt swift\_symba7} that is capable of correctly handling 
the closely-packed planetary systems (\citet{Levisonetal11}).} to handle close 
encounters and mutual collisions between planets. 
\par
The paper is organized as follows. Section 2 explains the set-up of our numerical 
simulations. The dynamics of growing planets is described in Section 3.  We then 
discuss the effects of the gas disk dispersal and the subsequent stage of purely 
$N$-body interaction of the remaining planets (Section 4). Conclusions are given 
in Section~5.
\section{Setup of numerical simulations}
\subsection{Disk parameters}
We used the hydro-dynamical two-dimensional code FARGO (\citet{Masset00}), in which 
the original $N$-body Runge-Kutta integrator was replaced (\citet{MorbyNesvo12}) with 
the symplectic integrator SyMBA (\citet{Duncan98}). The SyMBA code was specifically 
designed to handle close encounters and mutual collisions between planets.
As the hydro-dynamical simulations are CPU expensive, we were not able to run
many simulations to fully explore parameter space. Instead, we considered a few 
cases that illustrate different aspects of the problem.
\par
Two different disks were considered (denoted by $A$ and $B$ in the following) with 
the initial mass $M_A= 0.009M_{\odot}$ in case A and $M_B= 0.018M_{\odot}$ in case B,
where $M_{\odot}$ is the mass of the Sun. In each case we performed several simulations 
that differed in the prescription for the growth of the planets (see below). 
\par
We use units such that $G=1$ and $M_{\odot}=1$. The orbital period of a planet 
with semi-major axis $a=1$ is therefore $T=2\pi$. We normalize the time $t$ by 
$T$ in the following,  so that $t$ corresponds to the number of orbits at $a=1$, or years.  
The disk's kinematic viscosity coefficient is set to be 
$\nu = 10^{-5}$ in these units. { The initial surface density profile
scales with the distance $r$ from the star as $r^{(-1/2)}$.}
\par
Our computational domain consists of an annulus of the protoplanetary disk 
extending from $r_{min}$ to $r_{max}$. Different disk extensions have been used 
in different cases. In some cases (specified below), the disk had to be extended 
during the simulations when the migration caused the innermost planet to 
approach the inner boundary. To start with we used $r_{min}=0.5$, $r_{max}=4.5$, 
and a grid of $N_r=660$ { linearly spaced} radial cells and $N_s=700$ azimuthal cells. 

\par
The width of planet's horseshoe region is given, in the isothermal disk approximation 
(\citet{MassetDangelo06}), by:
\begin{equation}
x=\sqrt{(m/M_{\odot}) \over (H/r)}a \ .
\end{equation}
For example, for a planet of mass $m/M_{\odot}=3\times 10^{-5}$ (i.e. $10$ Earth masses)  
and disk aspect-ratio $H/r=0.05$ we get $x = 0.0245a$, where $a$ is the semi-major axis 
(see Section 2.2). The radial resolution of $0.006$ allows us to resolve the 
horseshoe region by at least 4 cells for any $a\geq 1$.
\par
{ The planetary contribution to the potential $\Phi$ acting on the disc 
is smoothed according to:
\begin{equation}
\Phi = -{Gm \over {\sqrt{d^2+\epsilon ^2}}}
\end{equation}
 where $d$ denotes the distance of a disc element to the planet
and $\epsilon$ is the smoothing-length. In our simulations we used
 $\epsilon= 0.5R_{H}$, where $R_H$ denotes the Hill radius. }
\subsection{Initial orbits}
Three 10-Earth-mass planetary cores were placed into the disk and were initially evolved 
till they reached a stable configuration in a resonance. We used a planet trap 
(\citet{Masset02,MassetMorby06,MorbyCrida08}) to halt the orbital migration of
the innermost core. \par The planet trap was set as a steep and locally positive 
surface-density gradient in the disk inside the initial orbital location of the 
innermost core. It allows the system of three cores to acquire stable, separated 
and non-migrating orbits. The planet trap is convenient way to mimic the situation
in real radiative disks where the non-isothermal effects can change the direction 
of the type-I migration (\citet{PaardeMelle06,KBK09}). The migration in the inner 
part of a real radiative disk can be directed outward, while it remains directed 
inward in the outer disk. This establishes the existence of a critical radius where
migration vanishes. The planetary cores migrating inwards will be collected near this
radius as in the case of a planet trap (\citet{Lyraetal10}).
\par 
The planet trap location was set at $a=3$ in case A and $a=2$ in case B. The local 
and positive surface-density gradient required to form the planet trap was created 
by imposing a transition in the viscosity from $4\nu$ to $\nu$ over $\Delta r = 1$ 
around the trap location (\citet{MassetMorby06})
The initial orbits of the three cores were chosen near the 5:4 resonant chain in
case A (semimajor axes $a_1=3.07$, $a_2 =3.62$ and $a_3=4.20$), and near the 3:2 
resonant chain in case B 
(semimajor axes  $a_1=2.1$, $a_2 =2.77$ and $a_3=3.66$). 
The initial eccentricities were set to zero. 
\par
In a first step, we followed the evolution of disk and cores, and waited till the cores 
arranged themselves in a stable resonant configuration. In case A, the cores 3 and 2 
ended up in the 6:5 resonance, and cores 2 and 1 in the 7:6 resonance. In case B, 
two cores reached a coorbital configuration (1:1 resonance) near the planet trap, and 
the third one ended up in the 6:5 resonance with the other two. The eccentricities 
remained small at this stage, $e\sim 0.01$-$0.02$, due to the strong damping of gas, 
and the orbits remained nearly coplanar.
\subsection{Mass growth}
Once the resonant configuration was achieved, the mass of each core was increased
from the initial value ($m(0)=3\times 10^{-5}M_{\odot}$) to one Jupiter mass 
($m_J=10^{-3}M_{\odot}$) as follows:
\begin{equation}
m(t) = m(0)+ (m_J-m(0))\sin^2({\pi \over 2}{(t-t(0))\over \Delta t})\ ,
\label{massgrowth}
\end{equation} 
where $t(0)$ was the time when the growth started, and $\Delta t$ was the growth
time interval. Different values of $t(0)$ were chosen for different planets, so 
that they grew in sequence. Sometimes, we let the innermost core grow first with
the other two growing later. In other cases we opted for growing the middle or 
outer core first (see Section~3). 
\par
{ We didn't consider gas accretion  within the Roche lobe. This is a delicate point which is not well understood yet and which goes beyond the purpose of
the present work.}\par The criterion for collision is that the distance between planets becomes equal or lower to one Jupiter radius.\par
The timescales for the planetary growth that we adopt ($\Delta t =10^{3}$, 
$3\times10^{3}$, and $4\times 10^{3}$, see table \ref{tab:1}) are too short 
to be realistic. These timescales are dictated by our current CPU power (using the 
parallel FARGO code and 30 CPUs we compute 10 to 100 orbits in 1 hour, depending 
on the disk parameters). Slower growth rates will be investigated in the 
future.
\par
When a planet grows to Jupiter mass, it is expected that the planet trap should 
become ineffective and the planet should start migrating inward. We find this 
behavior in our simulations.
\par 
\section{Dynamics in the gas disk}
\par
Here we discuss the orbital dynamics of growing planets in the full gas disk.
Each system is evolved over $20000$ to $65000$ years after the growth of the 
last planet. To identify the different settings of our simulations we label 
them $A_{i,j,k}$ for case A and $B_{i,j,k}$ for case B, where indices $i,j,k \in  [1,3]$  
indicate the growth order of the three cores (initially $a_1\leq a_2 <a_3$). 
The characteristics of each simulation are reported in Table \ref{tab:1}.
\subsection{Case $A_{2,1,3}$}
%


\begin{table*}
\begin{center}
\begin{minipage}{150mm}
\begin{tabular}{|llllr|}
\hline\noalign{\smallskip}
Simulation  & $[r_{min},r_{max}]$ & $\Delta t$ & ($N_r$,$N_s$) &  \\
\noalign{\smallskip}\hline\noalign{\smallskip}
\hline\noalign{\smallskip}
$A_{2,1,3}$ & $[0.5,6.5]$  &   $10^{3}$ &  (660,700) & \\
$A_{3,1,2}$  & $[0.5,9.5]$ &  $\Delta t _3=4\times10^{3}$ & (660,700) & \\
           &              & $\Delta t _{1,2}=10^{3}$  & \\ 
$A_{3,2,1}$ & $[1,9.5]$    &     $3\times10^{3}$ & (660,700) &  \\
          & $[0.5,9.5]$  $t \geq 20000$ & & (700,700) & \\
$B_{1,2,3}$  & $ [0.5,4.5]$ &  $10^{3}$ & (660,700) & \\
           & $[0.1,4.5]$  $t \geq 8500$  &                    & (720,700) & \\
$B_{3,1,2}$  & $ [0.5,4.5]$ &  $10^{3}$ & (660,700) & \\
            & $[0.1,4.5]$  $t \geq 8500$  &         & (720,700) & \\
            & $[0.3,4.5]$ $t \geq 30000$ &  & (690,700) & \\
\noalign{\smallskip}\hline
\end{tabular}

\caption{Simulations parameters.}
\end{minipage}
\end{center}
\label{tab:1}       
\end{table*}

Figure \ref{trap3pl1} shows the results of simulation $A_{2,1,3}$.
We find that when core \#2 grows to Jupiter mass it starts migrating inward 
 and scatters the other two cores at $a>3$. { We remark that the initial location of the trap is at $a=3$, 
but the gap opened by core \#2 has shifted  the trap at $a=4$. Therefore, cores \#1 and \#3 remain trapped at quasi constant semi-major axis, respectively at $a\simeq 4$  and $a\simeq 5.6$ till core \#1  grows.}
As the fully grown planet \#1 starts migrating inwards at $t=8000$,
 { it opens a gap and shifts the trap location at $a\simeq 5.5$.}
Core \#3 is scattered out and remains near the new trap position at
$a\simeq 5.5$ till $t_3(0)=12000$.  Core \#3 then starts growing and inward migration begins. 
\par
Once all three planets reach Jupiter mass they migrate inward and 
evolve into mutual resonances. Their orbital eccentricities rapidly grow 
to large values by resonant interactions. The system becomes highly unstable. 
The gas density distribution is strongly perturbed at this point (Fig. \ref{gas150}) 
leading to complex gas-planet interactions. 
Finally, at $t=20000$ years, the inner planet is lost by plunging into the 
star.

Our criterion for collision with the star is that the pericenter of the planet's
orbit becomes smaller than $0.01$. The tidal effects are ignored in our simulation. 
In reality, however, the tidal effects should start to be dominant for  small 
pericenters, potentially leading to the circularization of the orbit in the hot 
Jupiter region (\citet{BeaugeNesvo12}). This effect would result in decoupling 
the planet from the other two. So, in any case, the inner planet's influence
on the other two is suppressed. The subsequent evolution of the two-planet 
system leads to stable  orbits with moderate eccentricities. 
\par
\subsection{Case $A_{3,1,2}$}
In our first simulation  core  \#3 was grown on $\Delta t = 10^{3}$. 
One of the two remaining cores was ejected 
from  the system during the growth of core  \#3.
We have therefore reconsidered this case with a longer phase of growth 
of core \#3, $\Delta t_3 =4\times 10^{3}$ (table \ref{tab:1}). In this case,  
the transition is less violent, core \#2 is scattered { out and its semi-major axis remains stable at the new trap location (after gap opening by planet
\#3), i.e. at $a\simeq 4.3$.} 
Core \#1 migrates inward at the same rate as the fully grown planet \#3 
(Fig. \ref{trap3pl2long}). The growth of core \#1 then leads to a phase when 
the other two planets are scattered outward. The subsequent dynamics is complex
with episodes of outward migration, and a rapid increase of the eccentricities 
after the growth of planet \#2. As in case $A_{2,1,3}$, the system becomes 
highly unstable. Planet \#1 is then ejected from the system. 
{ These early ejections could be related to free-floating planets (\cite{Sumietal11}).}

The orbits of the
remaining two planets are chaotic  till $t=52594$, when the two 
planets merge.\footnote{Note that merging events may happen too often in our
simulations due to the coplanar approximation of the system.}
The remaining giant planets migrates inward and converges to a circular orbit. 
\par
\subsection{Case $A_{3,2,1}$}
In case $A_{3,2,1}$, all the three cores grow on $\Delta t = 3\times 10^{3}$
(Fig. \ref{trap3pl2pl1}). The growth phase leads to the inward migration 
of the growing planet \#3. The other two cores are scattered outward.
The subsequent phase  is quite different with respect to the two previous 
cases. Here the system settles in a stable resonant 2:1 configuration with 
planets slowly migrating inward, and eccentricities reaching moderate values
($e\sim0.2$-0.3).\par
{  We remark that three of the four planets found around the red dwarf
Gliese 876 are on the  triple 2:1 resonant configuration (\cite{Rivera10}).  The masses of the planets as well as their distance from the star are different from our simulation so that our comparison is only qualitative but nevertheless interesting.} \par
 To avoid spurious boundary effects, the disk was extended 
to $r_{min}=0.5$ at $t=20000$, when the pericenter of planet \#3 was close to $1$.   
The radial density profile has been extrapolated from the inner disk edge, 
and the simulation restarted with the extended disk.

\par 

\begin{figure}
\includegraphics*[height=7.0truecm,width=7.5truecm]{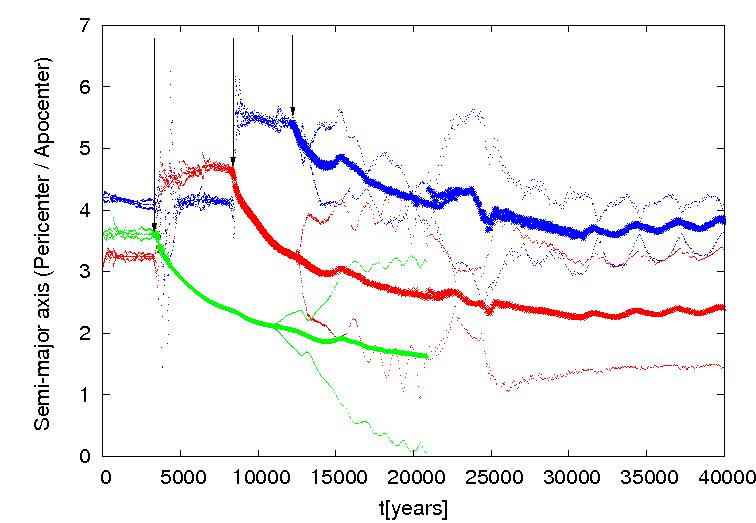}
\caption{Dynamical evolution of the planets in simulation $A_{2,1,3}$.
Each planet is represented by different color: red for planet \#1 (initially the 
innermost one); green for planet \#2 (middle); blue for planet \#3 (outermost). For 
each planet the three curves denote the pericenter, semimajor axis and apocenter 
as a function of time. The masses of the cores grow to Jupiter mass according to 
Eq. \ref{massgrowth} on $\Delta t= 1000$ starting $t_2(0)=3200$, $t_1(0)=8000$ and 
$t_3(0)=12000$. The arrows highlight $t(0)$ for each planet.}
\label{trap3pl1}
\end{figure}

\begin{figure}
\vskip -0.5truecm
\hskip -0.8truecm
\includegraphics*[height=8.6truecm,width=8.0truecm]{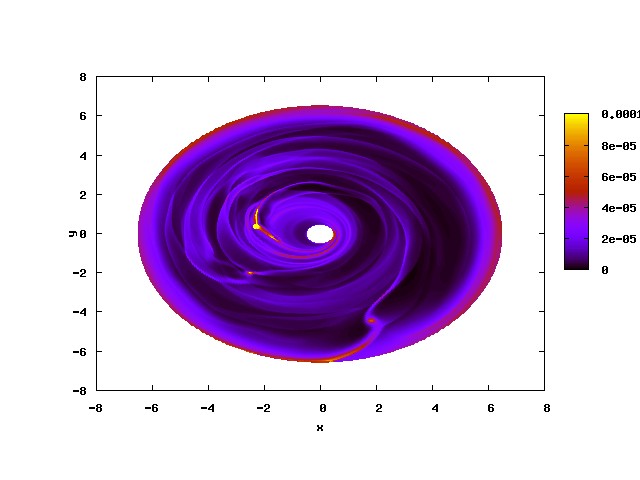}
\caption{Gas surface density in the simulation shown in Fig. 1 at $t=15000$. Three 
eccentric Jupiter-mass planets produce a complex distribution of gas.  Color scale
range was chosen such that dark blue corresponds to values $\leq 10^{-5}$ and yellow 
to values $\geq 10^{-4}$ (in dimensionless units).}
\label{gas150}
\end{figure}

\begin{figure}
\includegraphics*[height=7.0truecm,width=7.5truecm]{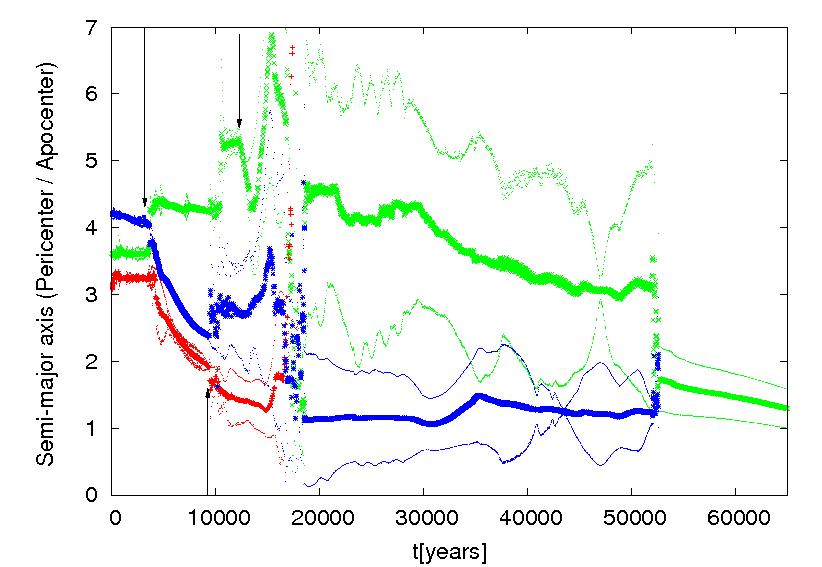}
\caption{The same as Fig. \ref{trap3pl1} for case $A_{3,1,2}$. 
The arrows highlight the $t(0)$  values: $t_3(0)=3200$, $t_1(0)=8700$ 
and $t_2(0)=12000$.}
\label{trap3pl2long}
\end{figure}

\begin{figure}
\includegraphics*[height=7.0truecm,width=7.5truecm]{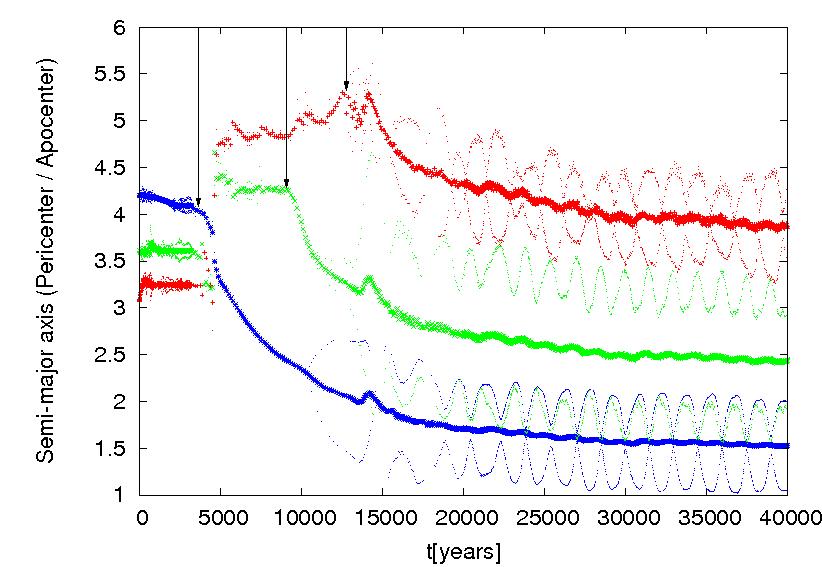}
\caption{The same as Fig. \ref{trap3pl1} for case  $A_{3,2,1}$. The growth of 
cores started at $t_3(0)=3200$, $t_2(0)=8600$ and $t_1(0)=12000$, as indicated 
by the arrows.}
\label{trap3pl2pl1}
\end{figure}


\subsection{Case $B_{1,2,3}$}
\begin{figure}
\includegraphics*[height=7.0truecm,width=7.5truecm]{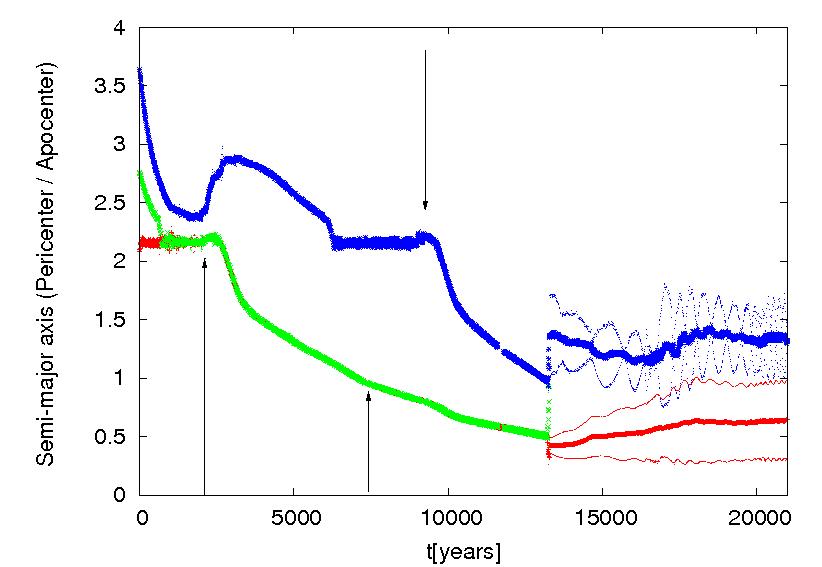}
\caption{The same as Fig. \ref{trap3pl1} for case  $B_{1,2,3}$. The growth of cores 
started at $t_1(0)=2000$, $t_2(0)=7000$ and $t_3(0)=9000$, as indicated by the arrows.}
\label{trap2pl1}
\end{figure}

In this case, cores \#1 and \#2 become coorbital at the trap location. It is interesting 
that they remain coorbital during the growth phase even if they do not grow at the same 
time (see Fig. \ref{trap2pl1}). The third core is scattered out when planet \#1 grows. 
It then migrates inward until it reaches the trap at $a\simeq 2$. When its mass starts 
increasing, \#3 migrates inwards, and all three planets reach in a compact orbital 
configuration. The disk has been extended to $r_{min}=0.1$ at $t=8500$ using the same procedure 
as for case $A_{3,2,1}$. The extended disk is followed with a time-step of $\simeq10^{-3}$.
To calculate 10 orbits this requires about 1 hour on 30 CPUs. The two coorbital planets 
merge at $t=13275$ and the third one is scattered out. The eccentricities of the two 
remaining planets grow to moderate values. The planets end up in the 3:1 resonance.

\subsection{Case $B_{3,1,2}$}
\begin{figure}
\includegraphics*[height=7.0truecm,width=7.5truecm]{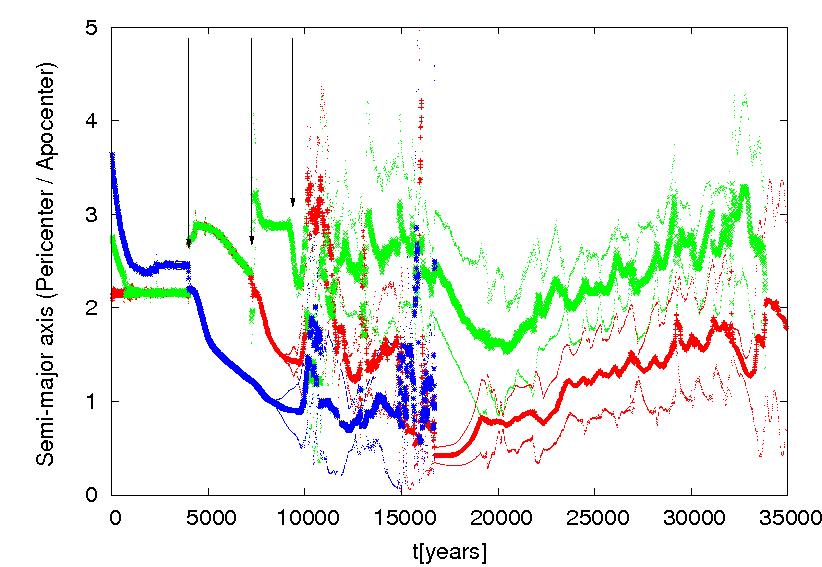}
\caption{The same as Fig. \ref{trap3pl1} for case case $B_{3,1,2}$. The growth of 
cores started at $t_3(0)=3700$, $t_1(0)=7000$ and $t_2(0)=9000$, as indicated by the 
arrows.}
\label{trap2pl3}
\end{figure}

Core \#3 grows first and scatters two coorbital cores outward.
{ The two cores appear to be on the trap at very small angular separation
$\Delta \alpha$ with: $10^\circ < \Delta \alpha < 30^\circ $. It is therefore possible that the scattering event affect them on the same way. Actually, as a result of the scattering they   remain coorbital and only their angular separation changes drastically.
The two orbits separate at $t=7000$ when core \#1 starts growing.}  Once that happens, core \#1 scatters 
the core \#2 outward (Fig. \ref{trap2pl3}). After the growth of core \#2, a complex instability 
arises resulting in the ejection of \#3 from the system. The two remaining planets have 
moderate eccentricities and persist on chaotic orbits { showing an outward migration trend till $t=33900$ when the two planets merge.  This case is comparable to case $A_{3,1,2}$. \par
The disk has been extended to $r_{min}=0.1$ at $t=8500$. In order to  follow the chaotic evolution of the two planets on reasonable CPU times,  we have then reduced the disk domain increasing $r_{min}$ to $0.3$ at $t=30000$.}

\subsection{Summary}
From the limited number of cases that we have investigated so far, we can derive the
following tentative implications for the dynamics of systems with three giant 
planets and their interaction with a  gas disk:
\begin{enumerate}
\item If the gas lasts long enough after the growth of the last giant planet,
the system develops an instability and one planet is typically lost. 
 { These  ejections could be related to free-floating planets (\cite{Sumietal11})}.

Indeed, only in one case out of five we obtained a final stable system with three 
giant planets. This is at odds with the results of \citet{Matsumura10} who 
found that the three-planet systems often survive to the end of gas disk
lifetime. This difference may appear from the approximate treatment of the
gas disk in \citet{Matsumura10}. When the gas density is strongly perturbed 
as in Fig. \ref{gas150} it acts as an additional source of stochasticity in 
the planetary evolution. \citet{Marzari10}, who used a hydro-dynamical 
code similar to ours, also found that systems of fully-grown giant planets 
on close orbits {\it rarely} survive to the end gas disk lifetime.   
\item
The simplified two-planet systems, that emerge from the three-planet systems 
when the the third planet is eliminated, tend to be stabilized by their 
interaction with the gas disk. This was also pointed out in \citet{Marzari10,Matsumura10} and in \citet{MRA2008}.  { In some cases, the two-planet system shows chaotic 
evolution till the system is reduced by a merging event. Future work
on the full spatial problem will be needed to better explore the frequency of merging events.}
\item We did not find any case where the giant planets would end up on nearly-circular 
closely-packed orbits. This raises doubts about the applicability of the 
initial conditions used in the models of planet scattering after the gas 
disk dispersal (see Section 1; \citet{Chatterjee08,Raymond09,Juric08}).
\end{enumerate}

\section{Gas dispersal and gas-free dynamics}
In the previous section we assumed that the gas disk remains present after 
the accretion of the giant planets, that is until the planets reach a stable 
dynamical configuration. It is possible, however, that the gas dispersal 
occurred {\it during} the planetary instability or soon after it, such that
the planetary system did not have enough time to fully stabilize. Here 
we investigate these  cases. 
\par 
The gas density was reduced at each time step $dt$ as:
\begin{equation}
 \rho' =\rho(1-{dt\over \tau})\ ,
\label{dissi}
\end{equation}
where the coefficient $\tau\simeq 2000$ years.
This dissipation timescale is very short when compared  to the $10^5$ years
timescale considered for photoevaporation in \citet{Matsumura10}.  
We find that, if the gas is removed too fast,  planetary systems can become unstable.
In our simulations, we didn't observe any scattering events or merging during the
dissipation phase, so that we are confident that our results wouldn't change much using a longer dissipation timescale. \par
 We  recall that our purpose is not to quantitatively describe 
a specific phase of the giant-planet--disk interaction but to obtain a qualitative 
description of the whole phenomenon;  this justify also  the use of a dissipation function (\ref{dissi}) which is simple with respect to the 
description in \citet{Moeckel12}.

When the disk gas density drops to values below $\sim10^{-10}$ (in dimensionless 
units), the effect of gas becomes negligible and we continue the integration with
SyMBA (\citet{Duncan98}). The planetary systems are evolved for up to $10^{8}$ years.
\par

\begin{figure}

\includegraphics*[height=7.0truecm,width=7.5truecm]{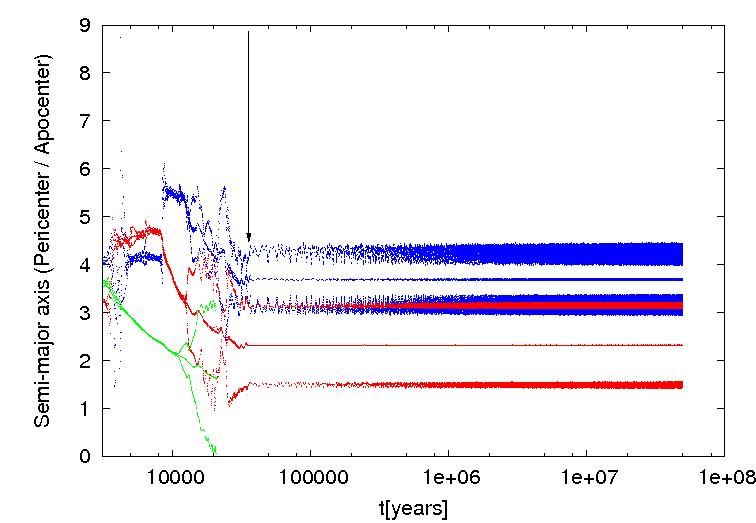}
\caption{Evolution of planetary orbits in case $A_{2,1,3}$.  The gas disk 
is removed at $[35000,37000]$, and an $N$-body integrator is used to follow
the gravitational interactions among planets for $t>37000$. The arrow 
indicates the beginning of the gas-free phase.}
\label{dissi35}
\end{figure}

We first tried a case where gas was removed after the planetary system has 
reached its final configuration. Fig. \ref{dissi35} shows the orbits for case 
$A_{2,1,3}$, where the gas disk was removed in the time interval $[35000,37000]$. 
During the removal, planetary migration slows down and the two remaining 
planets stay in the 2:1 resonance for the whole $N$-body integration. The same 
applies to $B_{1,2,3}$, where no scattering event was found after the gas 
dispersal (gas was removed at $[19000,20000]$ in this case).
\par
\begin{figure}
\includegraphics*[height=7.0truecm,width=7.5truecm]{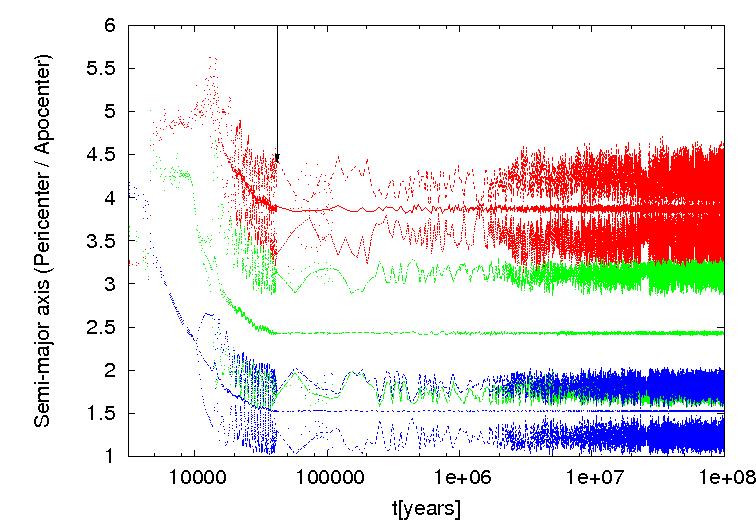}
\caption{Evolution of planetary orbits in case $A_{3,2,1}$.  The gas disk 
is removed at  $[39500,41000]$. The arrow indicates the beginning of the 
gas-free phase.}
\label{dissi30}
\end{figure}

In $A_{3,2,1}$, where three planets initially survived, they also survived for the 
whole length of the $N$-body integration. The three planets remained in 
the 2:1 resonant chain and no scattering among them occurred (Fig. \ref{dissi30}). 
This happened independently of the removal time (if chosen after $t=20000$ years
in Fig. \ref{trap3pl2pl1}) and independently of the removal timescale $\tau$. 
\par
In \citet{Matsumura10}, in agreement with our results, the two-planet systems also 
remained stable. For the three-planet systems, \citet{Matsumura10} found stability 
(e.g., see their Fig. 4) only in some cases. Unfortunately, having only one 
three-planet system we cannot test the statistical significance of our result. 
Note also that the previous versions of the SyMBA code used in \citet{Matsumura10}
had a later-identified problem when tracking closely-packed planetary systems (Levison 
et al. 2011). It has to be verified that this problem did not cause artificial 
instabilities in some of their integrations. 
\par
\begin{figure}
\includegraphics*[height=7.0truecm,width=7.5truecm]{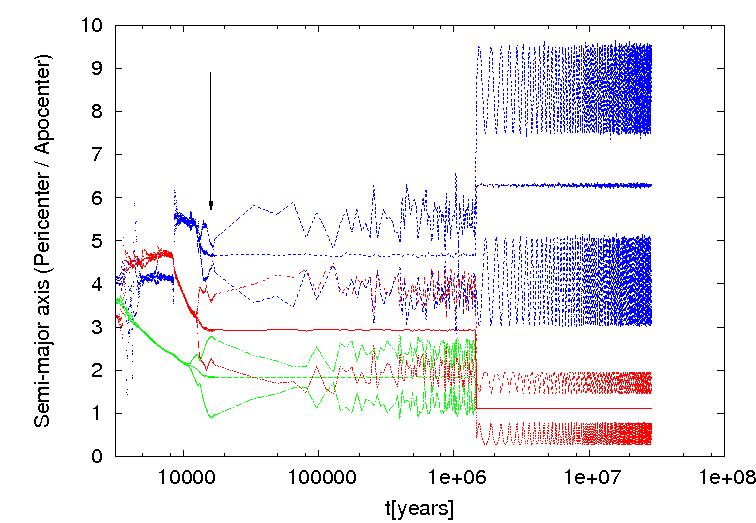}
\caption{Evolution of planetary orbits in case $A_{2,1,3}$. The gas disk 
is removed at  $[15000,17000]$. The arrow indicates the beginning of the 
gas-free phase.}
\label{dissi15}
\end{figure}

We now turn our attention to the possibility that the gas disk dispersed during the 
planetary-scattering phase. In Fig. \ref{dissi15}, we removed gas in the interval 
$[15000,17000]$ 
during the evolution of system $A_{2,1,3}$ (Fig. \ref{trap3pl1}).
In this case, the three giant planets undergo a gravitational scattering event in 
which one planet is ejected and the remaining two are sent onto highly-eccentric 
mutually-decoupled orbits (Fig. \ref{dissi15}). The same kind of evolution happened in all cases investigated here, provided that the gas disk was removed 
during the scattering phase.

\section{Conclusions}
We used a hydro-dynamical code to follow the orbital evolution of 
systems of three planets as they grew in sequence to Jupiter mass.
We found that the planet system changes drastically after the growth 
of each core. The orbital evolution of planets can be very complex. 
More often than not the orbits become unstable leading to a phase of 
planetary scattering. Planets can be ejected or merge (\citet{Marzari10}).  
\par
Once the system is reduced to two planets the dissipative effects 
of gas decrease orbital eccentricities of the remaining planets, 
and migrate planets into a new, stable resonant configuration. 
In only one case out of five, there was no instability happening after 
the growth of all three planets. The three-planet system remained 
in a resonant stable configuration in this case. 
\par
If the gas disk is removed after the new stable configuration is achieved, 
the orbital eccentricities remain low and the system is stable. This 
is at odds with the assumption typically made in the planet-scattering
models (\citet{Chatterjee08,Raymond09,Juric08}), where gas is 
ignored and the planets are initially placed on closely-packed unstable 
orbits. Here we show that these initial conditions may not naturally
arise from a previous stage, in which the planets interacted with 
their natal protoplanetary disk.   
\par
If the gas disk is removed during the planetary instability, planetary
scattering continues after the gas removal and the surviving planets 
can reach very eccentric orbits. However, given the short duration of 
the planetary instability phase, the removal of gas during this phase 
would require special timing. For example, it may be possible that 
the giant planets generally form toward the end of the disk lifetime. 
Or, as long as there is gas in the system, the existing giant planets 
keep growing and new giant planets keep forming. This would lead to 
a richer sequence of planetary instabilities than the one investigated 
here. We will investigate this possibility in the future work. 
\par
Another possibility is that the number of giant planets that form in a 
typical disk is large ($>3$). The $N$-body simulations have already shown 
that the eccentricity distribution of exoplanets implies that 
at least three giant planet existed in a typical system after the gas 
disk dispersal.
Our results seem to suggest that, for this condition to be fulfilled, more 
than three planets have to form originally. 
\par
Alternatively, the giant-planet systems that emerge from gas disks 
are stable in isolation, as suggested by in the simulations performed 
in this work, but become unstable due to external causes (interactions 
with smaller planets, effects of the planetesimal disks, etc.; 
e.g. Tsiganis et al. 2005, \citet{Levisonetal11}). 
\par
In conclusion, the results presented here show that the problem of 
understanding the dynamical paths leading to the surprisingly large 
eccentricities of extrasolar planets is not fully resolved. Future work 
should improve upon our efforts by using more realistic prescriptions 
for the planet growth and gas dispersal, extend the simulation to longer timescales, 
and perform a larger number of simulations so that the statistical 
significance of individual outcomes and their dependence on disk 
parameters is better understood.

\section*{Acknowledgments}
The computations have been done on the ``Mesocentre SIGAMM" machine, hosted 
by the Observatoire de la Cote d'Azur. D.N. acknowledges support from the NSF 
AAG program.

\label{lastpage}

\end{document}